\documentclass[preprint,12pt]{elsarticle}
\usepackage{amssymb}
\usepackage{amsmath}
\usepackage{xcolor}
\usepackage[a4paper,left=2.5cm,right=2cm,top=2.5cm,bottom=2.5cm]{geometry}
\newcommand{\ci}[1]{\cite{#1}}

\newcommand{\ba}{\begin{eqnarray}}
\newcommand{\ea}{\end{eqnarray}}
\newcommand{\be}{\begin{equation}}
\newcommand{\ee}{\end{equation}}

\newcommand{\beqs}{\begin{eqnarray}}
\newcommand{\eeqs}{\end{eqnarray}}

\begin{document}

\begin{frontmatter}

\title{TOTEM data and the real part of the hadron elastic amplitude \\ at 13 TeV}

\author{J.R. Cudell \&
O. V. Selyugin 
}
\address{ 
 \sl Li\`ege University, STAR Institute, Li\`ege, Belgium
\\
\sl  BLTPh, JINR, Dubna, Russia}

\begin{abstract}
We analyse the 13 TeV TOTEM data on elastic proton-proton scattering through a thorough statistical analysis, and obtain that
$\rho=0.096\pm 0.006$ and $\sigma_{tot}=(107.5\pm 1.5)$ mb. { Theoretical errors could lower the cross section by about 2 mb and increase $\rho$ by about 0.002. We also show that these results do not imply the existence of
an odderon at $t=0$.
}
\end{abstract}

\end{frontmatter}

\section{Introduction}
The TOTEM collaboration has recently measured \ci{TOTEM1,TOTEM2} with great precision the hadron scattering amplitude in the very 
forward, low-$t$, region at an energy $\sqrt{s}=13$ TeV. These data show some tension with the best fit of the COMPETE 
collaboration 
\ci{COMPETE}
and this has lead   to the claim that the odderon \ci{oddinv} has been discovered at zero momentum transfer, $t=0$ 
\ci{TOTEM2,nicodd,CERNodd}. As the odderon
has probably been seen at non-zero $t$ at the ISR \cite{Break}, this would solve one of the puzzles of hadronic physics, as so far no 
trace of it has been 
found at $t=0$. It is thus important to independently check the analysis of Ref. \ci{TOTEM2}.

Before doing this, we want to note that the central value of the COMPETE fits  (which agrees with the TOTEM 
cross section measurements) has no special significance. COMPETE evaluated the systematic error and produced 
a band of possibilities for  models without odderon exchange. The actual predictions, for $\sqrt{s}=13$ TeV, were 
86 mb $\leq \sigma_{tot}\leq $ 117 mb for the total cross section and 
0.058$\leq\rho(t=0)\leq$0.145 for the ratio of the real part of the amplitude to its imaginary part.  This shows that the LHC data 
do constrain the fits, and that the best fit of 16 years ago is no longer favoured. This does not mean that no fit can accommodate the 
new data without an odderon.

This letter will first describe the theoretical ingredients used in the present analysis, and show our best fit. We shall then explain
what the allowed ranges of forward hadronic observables are in the light of \cite{TOTEM2}. We shall also note the importance of the 
data of \cite{TOTEM1}, and stress the role of the normalisation factor in the data.

\section{Theory}
At extremely small values of $|t|$, the proton-proton elastic cross section is described by Rutherford scattering, i.e. the QED amplitude. 
At values of $|t|$ larger than 0.01 GeV$^2$, it is dominated by the purely hadronic amplitude, which comes from pomeron (and maybe 
odderon) exchange(s) at high energy. In between, both amplitudes matter, and the phase of the hadronic amplitude can be deduced 
from the interference between the photon-exchange amplitude 
 and the pomeron-exchange amplitude. 
\newcommand{\F}{{\cal A}}

Hence one needs to calculate the interference between the known amplitude due to photon exchange
and an unknown one due to pomeron exchange(s).  
The differential elastic cross section is described by the square of the elastic scattering amplitude $\F(s,t)$ divided by $s$.

The complete amplitude includes the electromagnetic (Coulomb) and hadronic (Nuclear) interactions and can be expressed as
\ba
  \F(s,t) =\F_{C}(s,t) \exp{(i \alpha \varphi (s,t))} + \F_{N}(s,t) .
\ea

The Coulomb amplitude can be conveniently split into several non-inter\-fe\-ring parts, according to the helicities of
the initial and of the final states. In the one-photon-exchange approximation, one gets five independent amplitudes{, which must be summed to obtain $\F_C$} \ci{gaz,pred}:
\begin{eqnarray}
 \F_C^{++\to ++}(s,t) &=& \F_C^{+-\to +-}(s,t)= \frac{s\alpha}{t} \ F_1^2(t)^2, \nonumber \\
 \F_C^{++\to --}(s,t)&=&-\F_C^{+-\to -+}(s,t) = s\alpha \ F_2^2(t), \nonumber \\
 \F_C^{++\to +-}(s,t) &=& - \frac{s\alpha}{ \sqrt{|t|}} \ F_1(t) \  F_2(t),
\end{eqnarray}
 with the Dirac, Pauli and dipole form factors given by
\ba
F_1(t)&=& \frac{ 4 \ m_p^2 - \mu_p  \ t}{4 \ m_p^2 - t} \ G_D, \nonumber \\
F_2(t)&=&\frac{4 \ m_p^2 \ (\mu_p - 1) }{4 \ m_p^2 - t} \ G_D, \nonumber \\
G_D(t)& = &  \frac{1}{(1-t/0.71)^2}, \ \ \  \nonumber \\
  \label{emff}
\ea
and where $ \mu_p$ is the magnetic moment of the proton and $ m_p$ its mass.

 The  phase $\varphi(s,t)$ has been calculated and discussed  by many authors.
 For  high energies the first results were obtained by A.I. Akhiezer and I.Ya. Pomeranchuk \cite{akhi}
 for the diffraction on a black nucleus. Using the WKB approximation in potential theory, H. Bethe  \cite{bethe}
 derived it for proton-nucleus scattering and obtained
\ba
  \varphi = 2 \ln{\left({1.06\over a\sqrt{-t} }\right)},
   \label{bethe}
\ea
 where the parameter $a$ characterizes the range of the strong-interaction forces and was taken as the size of a nucleus.
 
  After some improvements \cite{solov,rix}, the most important result was obtained by  M.P. Locher \cite{loch} and then by
  B. West and D.R. Yennie \cite{wy}.
Through the calculation of the associated Feynman diagrams, they obtained
  a general expression    for  $\varphi_{CN}(s,t)$ in the case of point-like particles
  in terms of the hadron elastic scattering amplitude
 \ba
  \varphi(s,t) =  \log\left({s\over |t|}\right) - \int^{0}_{-s}
                  \frac{ d t^{'} }{ |t-t^{'}| } \  \
	       \left(1-\frac{  \F_{N}(s,t^{'})  }
                       {  \F_{N}(s,t)  } \right).
 \label{wy}
 \ea

  For an exponential $t$ dependence of the hadron amplitude
  $\F_{N} = h s \ \exp\left({B(s) t\over 2}\right)$, we can get
 \ba
  \varphi(s,t) = \mp \left[\log\left({B(s) |t|\over 2}\right) + \gamma\right],
  \label{wyph}
 \ea
 where 
  $\gamma$ is the Euler constant
  and the upper (lower) sign
  corresponds to the scattering of particles with the same (opposite)
  charges. We shall use this result to describe $\F_C$.
 
 For the hadron amplitude near $t=0$ and $\sqrt{s}=13$ TeV, we can use the simplest form
    \begin{eqnarray}
  \F_N(s,t)= s {\sigma_{tot}\over 4 \pi}  (i + \rho(t)) \  \exp\left({B(s) t\over 2}\right) .
\label{AN}
  \end{eqnarray}
  In our analysis, we follow TOTEM \ci{TOTEM2} and neglect the $t$-dependence of $\rho$.

{ The exponential form factor and the constant value of $\rho$ may seem simplistic assumptions, which are motivated only by the rather low number of parameters. Nevertheless, we shall see that they give an excellent description of the data at very small $|t|$, with however slightly different parameters from those of the TOTEM analysis.
We shall also explain that the form of the parametrisation matters in a very small interval of $t$, and hence the parameters should be considered as effective ones for those values of $t$. For completeness, we shall discuss the uncertainties associated with these assumptions at the end.} 

\section{TOTEM  13 TeV data}
TOTEM has published data for the differential elastic cross section at small $|t|$, from which they
derived $\sigma_{tot}=(110.6\pm 3.4)$ mb \cite{TOTEM1} and $\rho=0.10\pm 0.01$, with a theoretical error of 0.01 \cite{TOTEM2}.
The data have  systematic errors dominated by the uncertainty on the normalisation. So, in the following, we shall use the possibility to
change the normalisation of the data by a factor $n$, as explained in \cite{TOTEM2}.

We also consider three statistical models, all based on a $\chi^2$ statistics. The first one used the simple form
\begin{equation} \chi^2=\sum_i {(n d_i-\theta(t_i,\rho,\sigma_{tot},B))^2\over \sigma_i^2},\end{equation}
where $d_i$ is the central value of the data in the $i^{th}$ bin, $t_i$ is the preferred value of $t$ in the bin $ [t_i^{min},t_i^{max}]$, as 
given 
in \cite{TOTEM2} (note that $|t_i|\approx \sqrt{t_i^{min}t_i^{max}}$), $\theta(t_i,\rho,\sigma_{tot},B)$ is the 3-parameter 
theoretical model of $d\sigma/dt$, and $\sigma_i$ is the statistical error. The second statistics considered here is
\begin{equation} \chi^2=\sum_i {(n d_i-\theta(t_i,\rho,\sigma_{tot},B))^2\over \Sigma_i^2},\end{equation}
where $\Sigma_i$ is equal to the sum in quadrature of the statistical and  systematic errors.
Finally, the TOTEM collaboration has provided the correlation matrix $V$ of the systematic errors. Defining 
$W_{ij}=V_{ij}+\sigma_i\delta_{ij}$, one
can write 
\begin{equation} \chi^2=\sum_{ij} (n d_i-\theta(t_i,\rho,\sigma_{tot},B)) W^{-1} _{ij} (n d_j-\theta(t_j,\rho,\sigma_{tot},B)).
\label{chicorr}
\end{equation}
 \begin{table}[h]
\label{Table-1}
\begin{center}
\begin{tabular}{|c||c|c|c|c|} \hline
 79 points					&$\sigma_{tot}$ (mb) 	& $B$ (GeV$^{-2}$)	&  $\rho$  	&$ \chi^2$/d.o.f.    \\ \hline
 statistical  &	$111.75  \pm   0.05$&$20.77\pm  0.03	$ &	 $0.083\pm   0.004$	& 0.89      \\ \hline  
  statistical and systematic &	$111.75   \pm    0.05$&$20.77\pm 0.03	$ &	 $ 0.084\pm   0.004$	 &0.80       \\ \hline  
correlated &	$111.75   \pm    0.13$&$20.77\pm 0.05	$ &	 $ 0.085\pm   0.005$	 &0.89       \\ \hline 

\end{tabular}
 \caption{Fits to the  first 79 data points from TOTEM for $0.0008 < |t| < 0.07$), for a normalisation $n=1$. }\end{center}
  \end{table}
We consider the data at small $|t|$, as they determine the value of $\rho(0)$. Following \ci{TOTEM2}, we fit  the first 79 points at
$0.0008 < |t| < 0.07$. We show in Table 1 the results of this fit, for the three statistics considered here. The consideration of 
correlated errors is close
to using only statistical errors. This third statistics reproduces the central values of TOTEM\footnote{It 
agrees with the version 2 of the preprint \cite{TOTEM2}, whereas version 1 had a much smaller value for the $\chi^2/d.o.f.$}. TOTEM uses a much more complicated form \cite{kundrat} for the phase than Eq. (\ref{wyph}), but this agreement shows that the effect of the exact expression of the phase is minimal. The big 
difference is the size of the error bars, due to the fact that TOTEM in its parameter evaluation allowed a global shift in normalisation of 
5.5 \%, and incorporated this in the error estimate. This brings us to the central question concerning these data. How well do we know 
their normalisation? And how should one take this uncertainty into account?

The first thing one can do is repeat the exercise of Table 1, but allow the data to be globally shifted by a factor $n$. As this factor 
controls the total elastic 
cross section measured by TOTEM in \ci{TOTEM1}, we use the value $n=1.000\pm 0.055$, which corresponds to their measurement 
$\sigma_{el}=(31.0\pm 1.8)$ mb, and treat the error as a statistical error.
We see in Table 2  that the normalisation factor prefers to be around 0.9, so that the total cross section is correspondingly lowered by 
about $1 \sigma$.
This is reminiscent of the situation at lower energies \cite{SC16}, where we also found that the data of TOTEM needed to be lowered 
by about 10~\%. 

 \begin{table}
\label{Table-2}
\begin{center}
\begin{tabular}{|c||c|c|c|c|c|} \hline
 79 points 					&$\sigma_{tot}$ (mb) 	& $B$ (GeV$^{-2}$)	&  $\rho$  	& $n$ 		&   $ \chi^2$/d.o.f    \\ \hline  
 statistical  &	$107.2 \pm       1.9$&$20.81\pm  0.04	$ &	 $ 0.098\pm   0.008$	& $0.92\pm 0.03$&0.82      \\ \hline  
  stat. and sys. &	$107.2 \pm       2.1$&$20.81\pm  0.04	$ &	 $ 0.098\pm   0.008$	& $0.92\pm 0.04$&0.75      \\ \hline  
correlated &	$106.4   \pm    2.2$&$20.80\pm 0.06	$ &	 $ 0.098\pm   0.008$	& $0.91\pm 0.04$&0.81      \\ \hline    					 
\end{tabular}
 \caption{Fits to the  first 79 data points from TOTEM for $0.0008 < |t| < 0.07$.{ \protect\footnotemark 
 }}\end{center}
  \end{table}   
   \begin{figure} [h]    
  \includegraphics[scale=0.6]{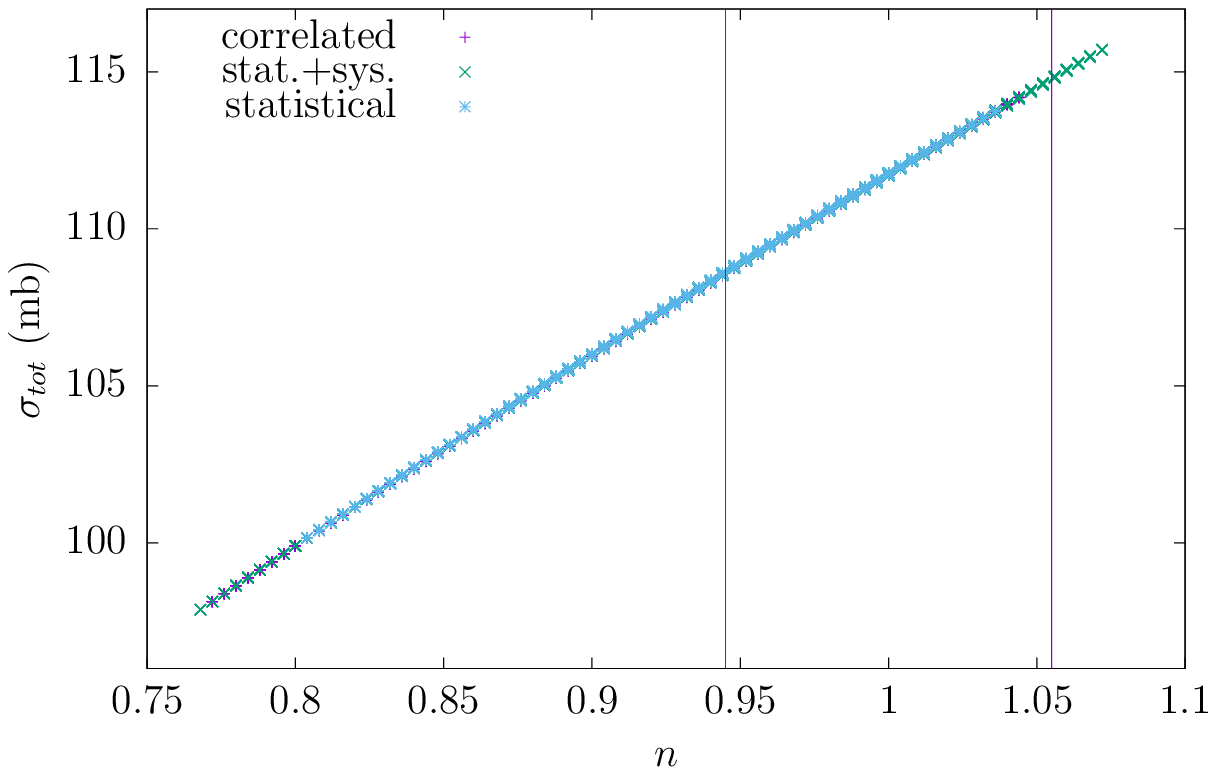}\includegraphics[scale=0.6]{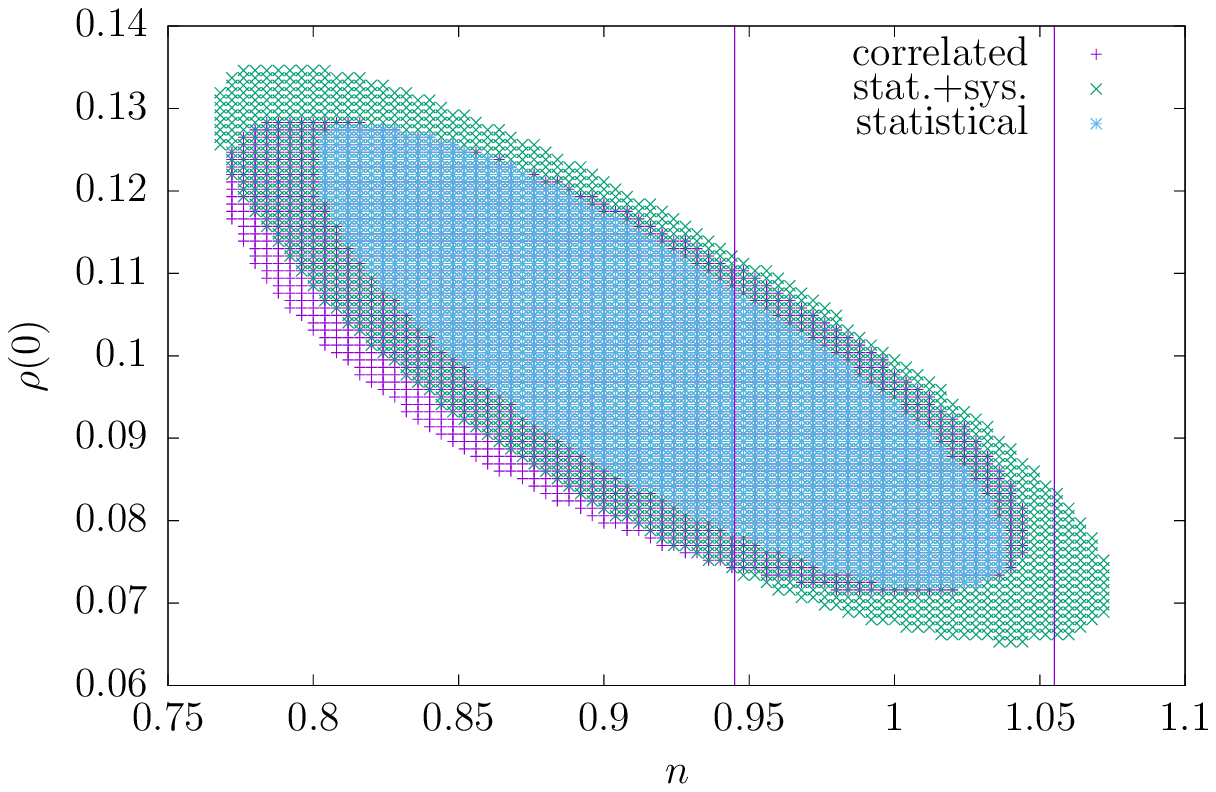}\\
   \includegraphics[scale=0.6]{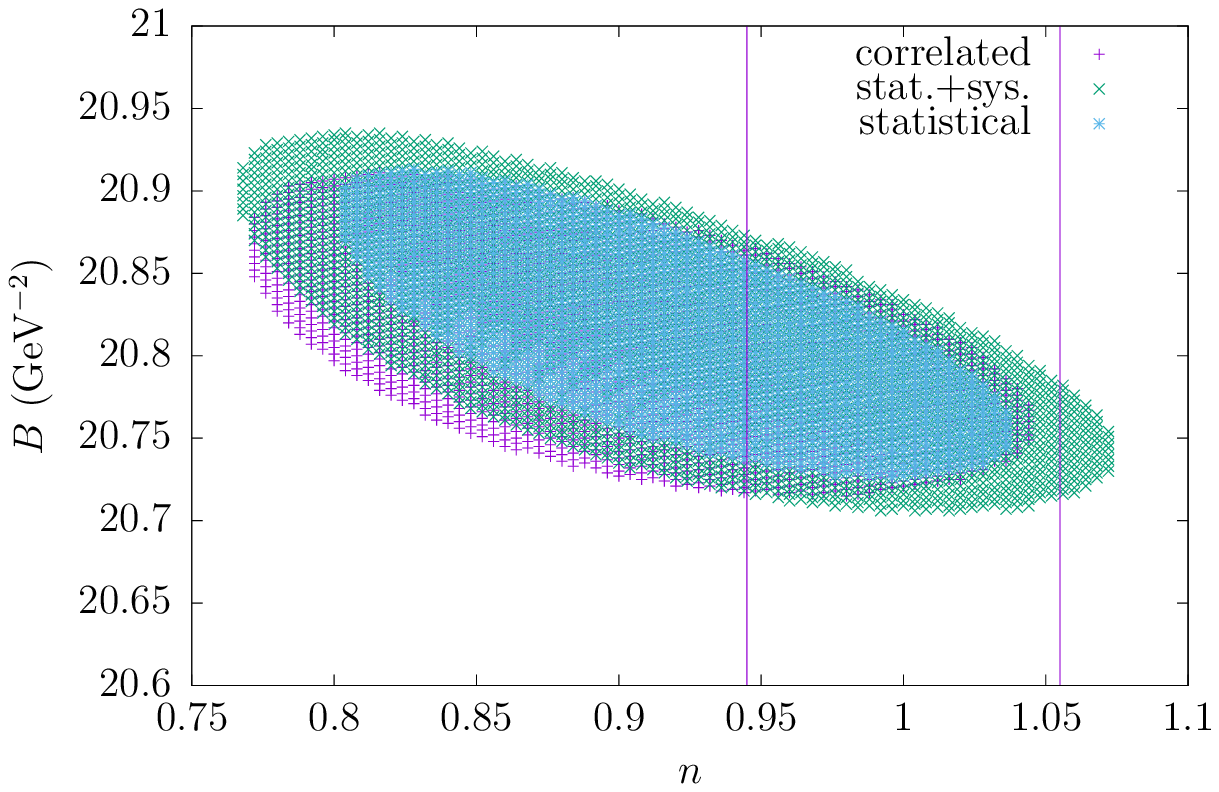}\includegraphics[scale=0.6]{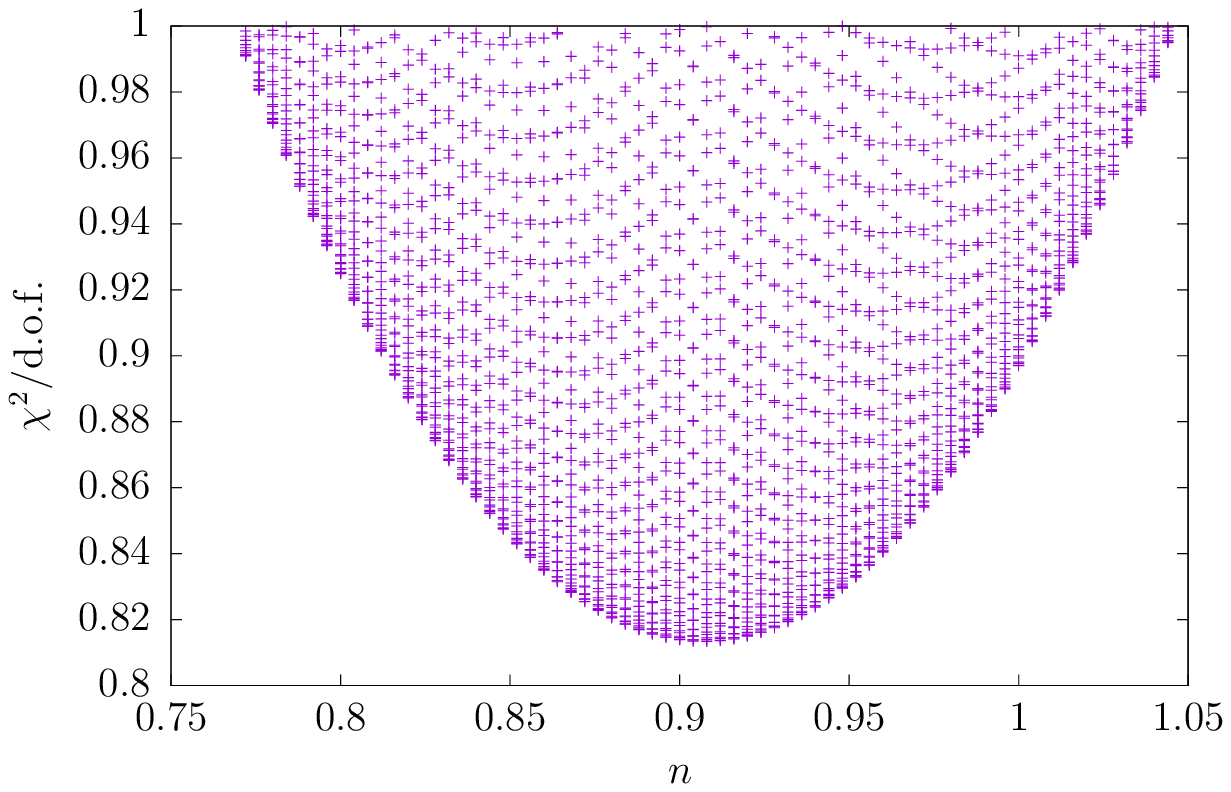}
 \caption{Regions for which $\chi^2/d.o.f$ is smaller than 1, for the three statitics used in this letter. The vertical
 lines indicate the regions allowed in the TOTEM analysis of \cite{TOTEM2}.}     \label{Regions}  
  \end{figure}  
  \footnotetext{ The previous determination of the normalisation \cite{TOTEM1}, i.e. $n=1.000\pm 0.055$, is treated as a data point and included in the $\chi^2$.} 
 All the results shown so far correspond to the usual definition of errors, i.e. the minimum $\chi^2$ increases by 1 unit if a parameter is 
 moved by 1 $ 
 \sigma$. This definition makes sense for fits with $\chi^2$/d.o.f. $\approx 1$. In the present case, the values of $\chi^2$ may be too 
 low to use this
 to determine the parameters. One could for instance allow all the parameter values such that $\chi^2$/d.o.f. $
 \leq 1$. Hence in Fig.˜1, we show the regions with $\chi^2/d.o.f$ less than 1 for $\sigma_{tot}$, $\rho$ and $B(0)$ as functions of the 
 normalisation $n$, for the three statistics used in this letter, together with the values of $\chi^2/d.o.f.$ of all the points in these 
 regions, this time only for  the correlated $\chi^2$. 
 
 Clearly, the question of the correct value of $n$ is of great impact on the results. We shall proceed with a complementary analysis to 
 settle this question  \cite{astro}.
  
\section{A Bayesian analysis}  
The data of TOTEM \cite{TOTEM1} can be translated into two prior probability densities, for $\sigma_{tot}$ and $n$, which we assume 
gaussian:
\begin{eqnarray}
\mathcal{G}_1&=&{\cal N}_1 \exp\left(-\frac {(\sigma_{tot}-110.6\ \mathrm{mb})^2}{2 (3.4\ \mathrm{mb})^2} \right)\\
\mathcal{G}_2&=&{\cal N}_2 \exp\left(-\frac {(n-1.000)^2}{2 (0.055)^2} \right)
\end{eqnarray}
with ${\cal N}_1$ and ${\cal N}_2$ two normalisation so that these probability densities integrate to 1.

We can then define a probability density for the new data
\begin{equation}
\mathcal{L}={\cal N} \exp\left(-\frac {\chi^2(x_k|d_i)}{2} \right){\cal G}_1{\cal G}_2
\end{equation}
where $x_k$ are our 4 parameters $x_1=\sigma_{tot}$, $x_2=\rho$, $x_3=B$ and $x_4=n$,
and $\chi^2$ is the correlated statistics of Eq. (\ref{chicorr}),
as in ref.˜ \cite{TOTEM2}.
 
\begin{figure} [h]
 \centerline{ \includegraphics[scale=0.5]{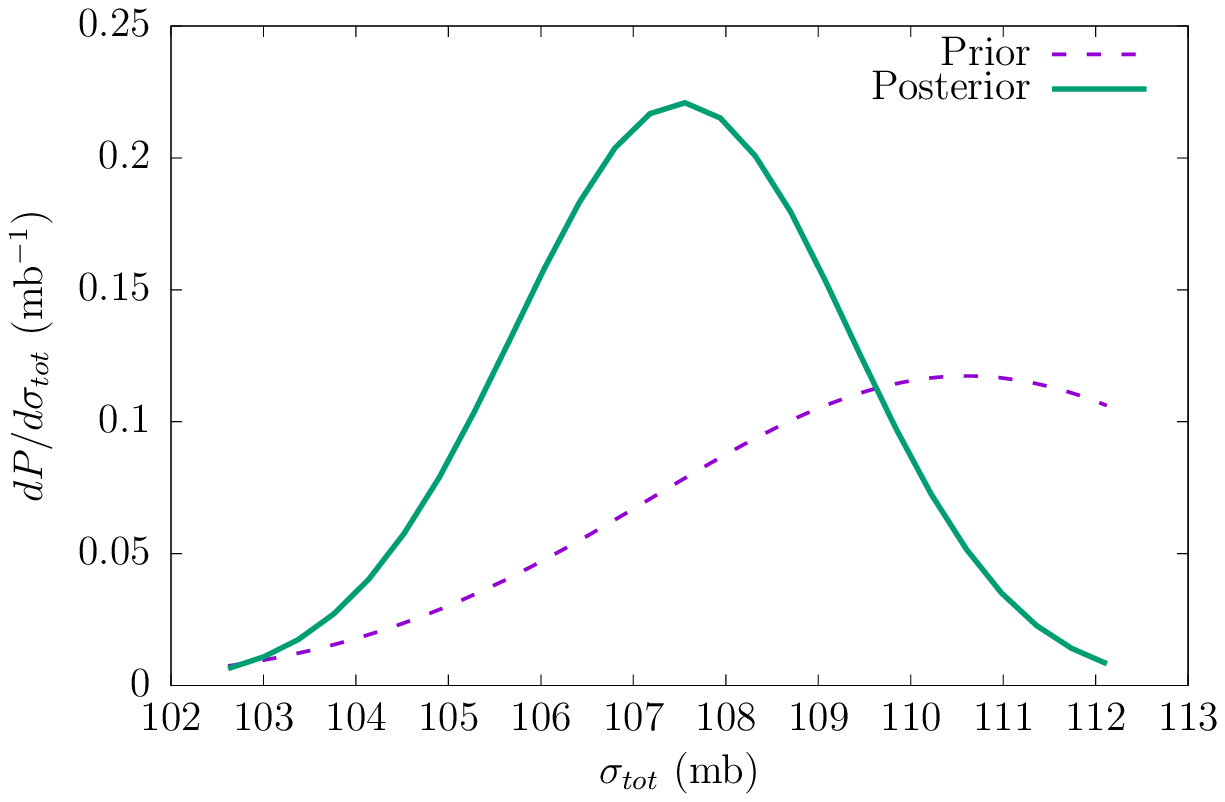} \includegraphics[scale=0.5]{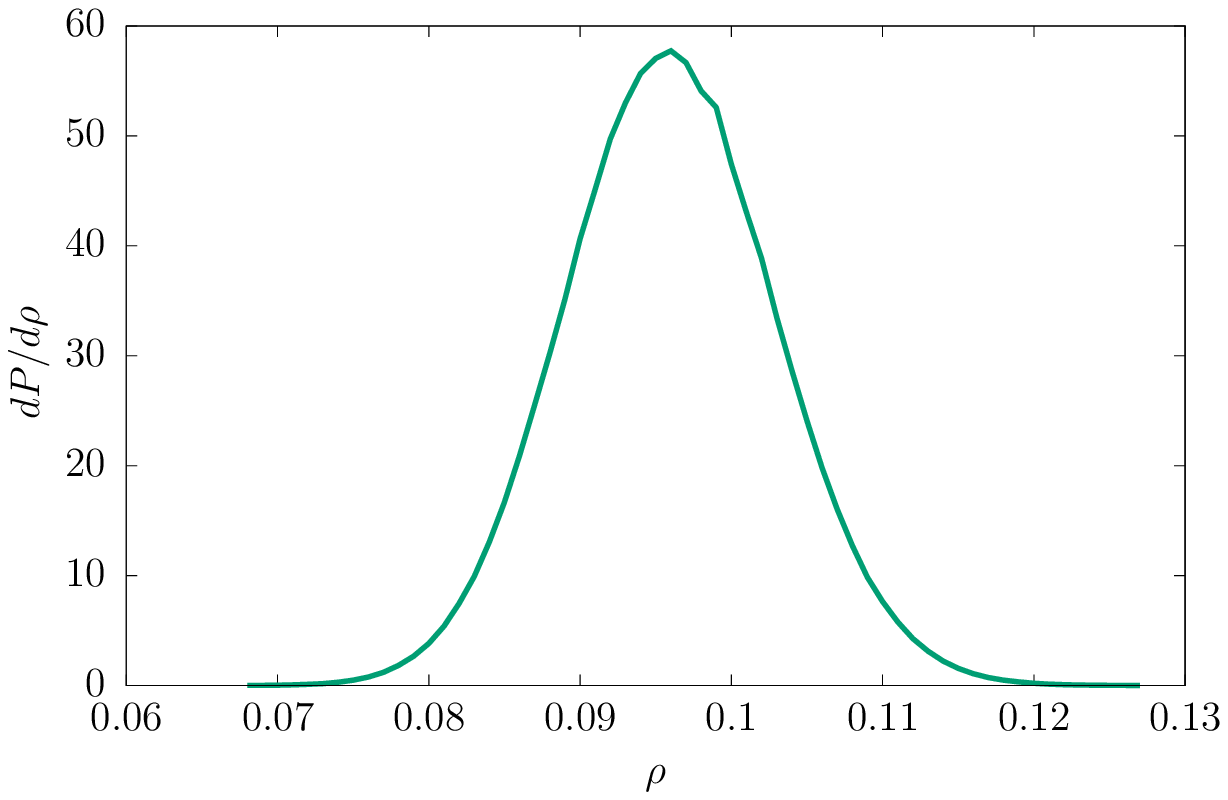}}
 \centerline{ \includegraphics[scale=0.5]{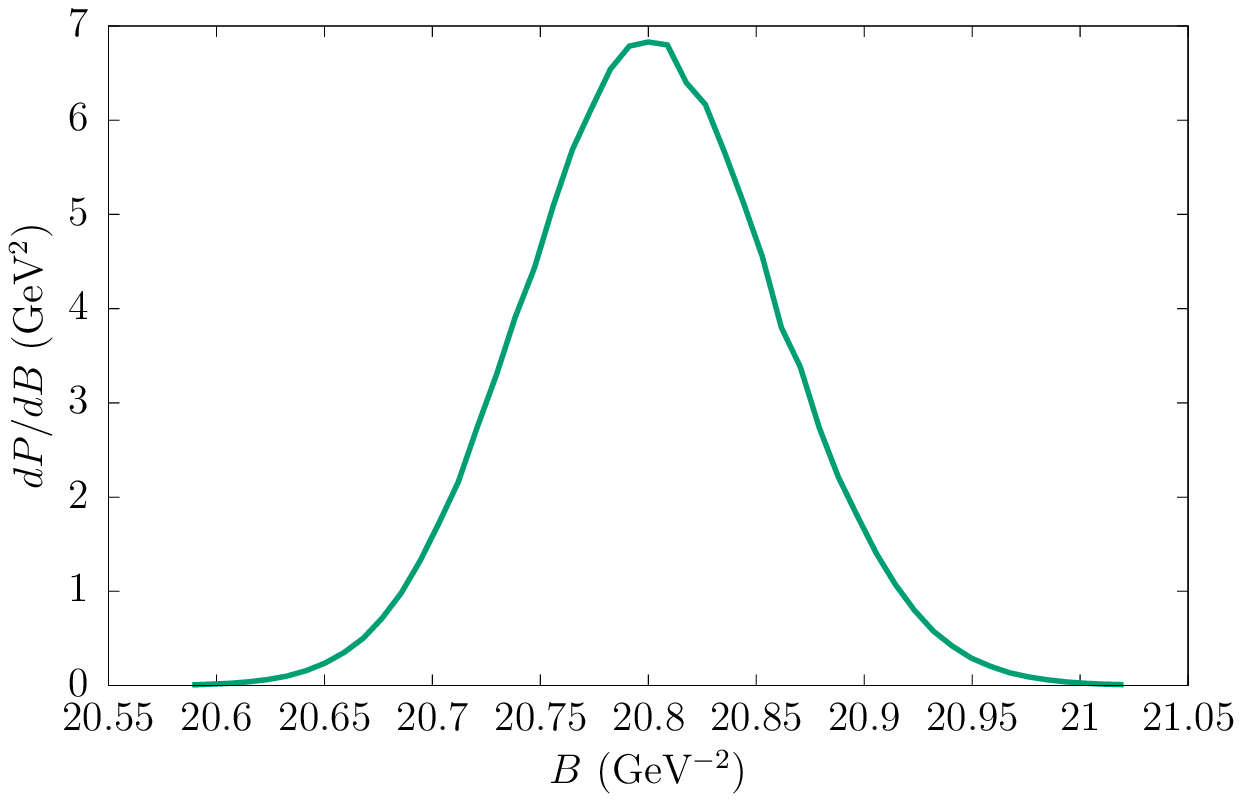} \includegraphics[scale=0.5]{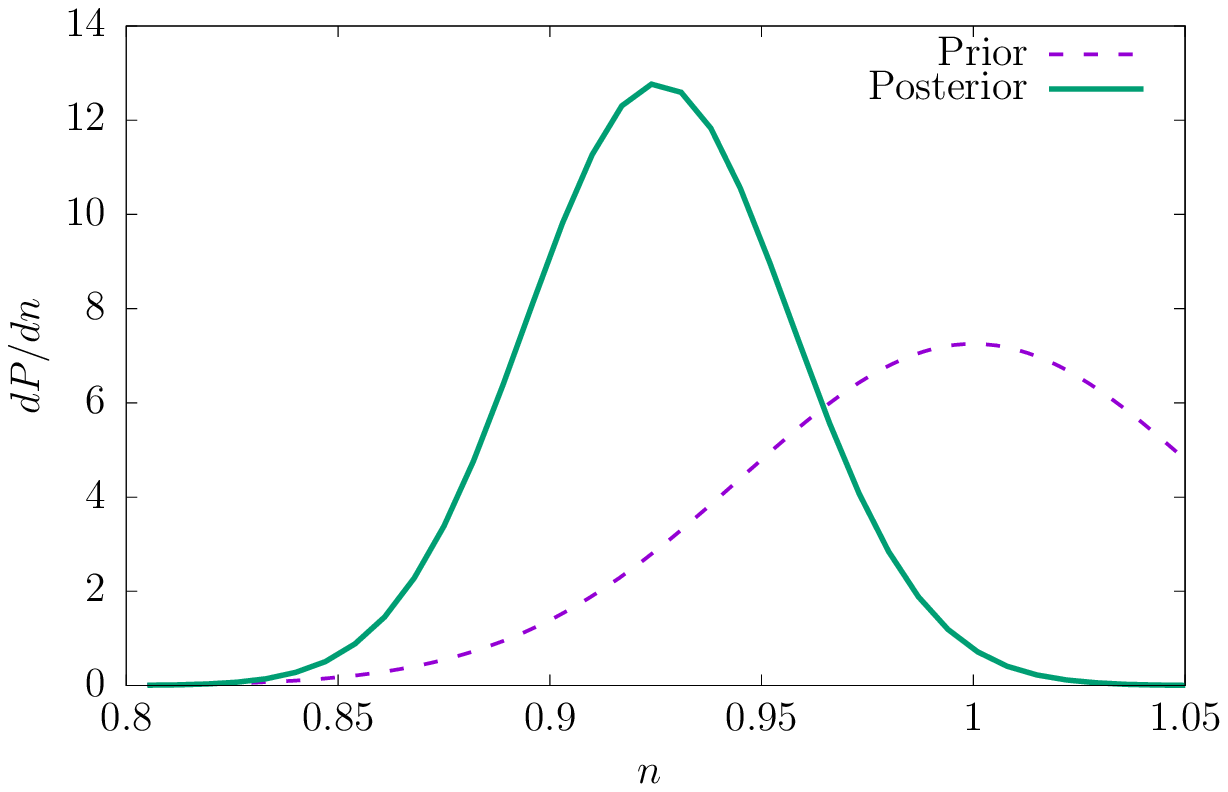}}
  \caption{Posterior probabilities of the parameters.}     \label{Bayes}
 \end{figure}

 $\cal N$ is defined so that
\begin{equation}
\left[\prod_{k=1,2,3,4} \int dx_k\right]{\cal L}=1.
\end{equation}
We then consider the probability densities in which we integrate of all parameters but one:
\begin{equation}
{d{\cal P}\over dx_k}=\left[\prod_{j\neq k} \int dx_j\right]{\cal L}.
\end{equation}
We then obtain the curves of Fig. \ref{Bayes}. We see that the main conclusion of the previous analysis is confirmed: the total cross 
section is
lower than assumed by TOTEM. The other parameters take values similar to those of Table 2 as well, and we give their values in 
Table~3.

\begin{table}  [h]
\begin{center}
\begin{tabular}{|c|c|c|c|} \hline
Parameter & Central value   & $1\sigma$ error & $2\sigma$ error \\ \hline
$\sigma_{tot}$ (mb)&107.5&	$\pm 1.5$&$\pm 2.7$\\ \hline   
 $\rho$&0.096	&$\pm 0.006$&{$\substack{+0.012 \\ -0.011}$} \\ \hline        
  $B$ (GeV$^{-2}$)&20.800&	$\pm 0.053$&$\pm 0.097$\\ \hline        
   $n$&0.924	&$\pm 0.028$&{$\substack{+0.041 \\ -0.049}$}\\ \hline        
\end{tabular}
 \caption{Parameters and errors from the curves of Fig. \ref{Bayes}\label{Table-3}}
 \end{center}
\end{table}
{
\section{Theoretical error}
The total cross section and the $\rho$ parameter are derived quantities obtained via the 
differential elastic cross section, and as such they depend on the model used to describe 
elastic scattering \cite{CSPRL}. Now the first question is whether the simple form used in Eq.~(\ref{AN} )
provides a good description of the data. One of course has an excellent $\chi^2$, but the best
way to decide is to consider the residuals, i.e. the difference between the best fit and the data.
We show these in Fig.~\ref{fig.residuals}.
\begin{figure} [h!]
 \centerline{ \includegraphics[scale=0.7]{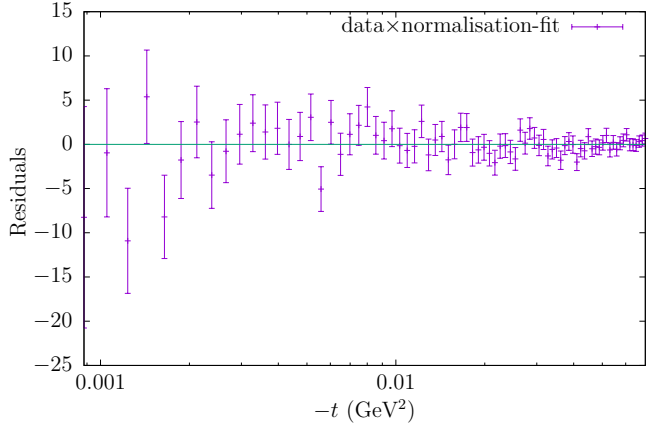} }
  \caption{Residuals corresponding to the correlated fit of Table 2.}  
  \label{fig.residuals}
 \end{figure}
Clearly, there is no significant departure from an exponential, and the residuals are random.
So, in this limited range of $t$, the assumption regarding the form factor matches the data.

However, the form factor enters in a more subtle way, as the phase formula involves integration
over the whole $t$ range, and different functional forms lead to different phases.
 In fact, formula (\ref{wyph}) is itself an approximation. We compare its results to
the exact form from \cite{SC16} in Table 3, and also consider the effect of the form factor. The errors are similar to those of Table 2, and the $\chi^2$ values are slightly lower, and won't be quoted here. 
We see that the form factor has the biggest effect on $\rho$, and that it can increase its value by 0.018.
\begin{table} [h]
\label{Table-4}
\begin{center}
\begin{tabular}{|c|c|c|c|c|} \hline
Form factor      & Phase&  $\sigma_{tot}$   & $\rho$    &  $n_i$      \\  \hline
exponential&Eq. (\ref{wyph})&	$106.4$&	 $ 0.098$	& $0.91$    \\
 exponential &   \cite{SC16}   &  $104.1$ & $0.106$    &  $0.87$      \\     
 dipole & \cite{SC16}   &  $106.3$ & $0.116$    &  $0.90$      \\\hline
\end{tabular}
\end{center}
 \caption{Dependence of the parameters
 on the formula for the CNI phase, and on the form factor, in the case of a constant $\rho$.
 } 
  \end{table}
 
 Another assumption in our analysis is that $\rho$ is constant.  This is clearly not the case in general \cite{CSPRL}. However,
 one has to realise that only a few points lead to a determination of $\rho$. We show separately  in Fig. \ref{2sigs} the contributions to the differential cross section of the hadron amplitude ${\mathcal A}_N$ and of the Coulomb amplitude ${\mathcal A}_C$. In the region where the 
 pomeron amplitude dominates, the effect of the real part is small, of order $\rho^2\approx 1\%$. 
 \begin{figure} [h!]
 \centerline{ \includegraphics[scale=0.7]{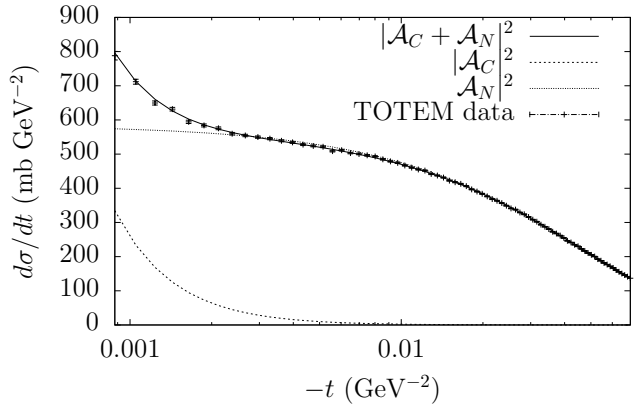} }
  \caption{Separate differential cross sections for the hadron contribution, for the Coulomb contribution, and for the total.}  
  \label{2sigs}
 \end{figure}
 There are a few points at the beginning (7 or 8) for which the Coulomb amplitude is not 
 negligible. For those points, at $0.001<|t|<0.003$ GeV$^2$, the interference term is present, and it
 is linear in $\rho$. It is thus this very small interval of momentum that contributes to the determination of $\rho$.
 Hence assuming a $t$ dependence for $\rho$ does not change the results much. For instance,
 assuming $\rho=\rho(0)(1+a t)$ leads to $\rho(0)=0.099\pm   0.008$ if $a<1$ GeV$^{-2}$.
 
 Hence the main source of theoretical error comes from the choice of form factor, and we estimate it at 2 mb on $\sigma_{tot}$ and 
 0.02 on $\rho$.}
\section{Conclusion}
We have shown that the data of \cite{TOTEM2} indicate that the measurement of the total cross section of \cite{TOTEM1} should be 
lowered by about 1 $\sigma$, and the elastic cross section by about 2 $\sigma$. This has already been pointed out at lower energies, 
using ATLAS and TOTEM results \cite{SC16}, but as we have shown here, it is true using only TOTEM data. 
\begin{figure} [h]
 \centerline{ \includegraphics[scale=0.35]{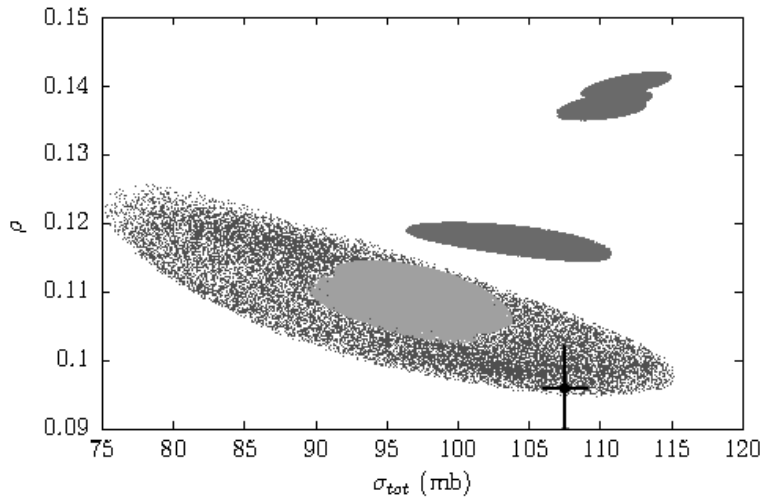} }
  \caption{Regions of the ($\sigma_{tot}$, $\rho$) plane allowed at the 1 $ \sigma$ level by the COMPETE fits. The upper three regions in dark gray
  correspond to fits of the form $Z+B\log^2(s/s_0)$, the light gray region to fits of the form $Z+B\log(s/s_0)$ and the dotted region to fits of 
  the form $A\log^2 s+B\log s+C$. The experimental point corresponds to the values of the present analysis.}\label{compete}
 \end{figure}
 
{ The second point concerns the need for an odderon. We could refit all existing data and check whether this is the case.
However, as it turns out, a parametrization published in 2002 \cite{COMPETE} agrees with the TOTEM data, without requiring an odderon. }
From the details of those fits \cite{web}, one can reconstruct the 
1 $\sigma$ allowed regions in the ($\sigma_{tot}$, $\rho$) plane. We show these regions in Fig.~5. Clearly, some of the 
parametrisations are ruled out. However, the two parametrisations for which $\sigma_{tot}=A\log^2 s+B\log s+C$ are allowed and fully
compatible with the data. It seems a refit would be a good idea, but the inference that the TOTEM data implies the existence of an 
odderon seems misguided.

In this letter, we concentrated on the simplest analysis of TOTEM data at 13 TeV. The main drawback is that we totally neglect the 
analytic properties of the hadronic amplitude, which constrain $\rho$ once the $s$ dependence of the amplitude is known, Such an 
analysis has however many problems. Even if one concentrates on high-energy data, there is disagreement between TeVatron 
experiments, and between LHC experiments, making the overall fit necessarily bad. Also, the analysis becomes model-dependent, as 
one does not know the true high-energy form of the hadronic amplitude. It might be possible to remove some of these problems by 
considering only TOTEM data, but we leave this to a future work.

\section*{Acknowledgements}{J.R.C. would like to acknowledge exchanges with P.V. Landshoff
and thank E. Martynov for his comments. O.V.S. would like to thank the University
of Li\`ege where part of this work was done.  This work was also supported by the
Fonds de la Recherche Scientifique-FNRS, Belgium, under grant No. 4.4501.15.}

\end{document}